\documentclass{ws-rv961x669}
\usepackage{ws-rv-van}     
\usepackage{amsfonts}
\usepackage{amssymb}
\usepackage{graphicx}
\usepackage{hyperref}

\newcommand\rf[1]{(\ref{eq:#1})}
\newcommand\lab[1]{\label{eq:#1}}
\newcommand\nonu{\nonumber}
\newcommand\br{\begin{eqnarray}}
\newcommand\er{\end{eqnarray}}
\newcommand\be{\begin{equation}}
\newcommand\ee{\end{equation}}

\newcommand\lb{\lbrack}
\newcommand\rb{\rbrack}

\renewcommand\({\left(}
\renewcommand\){\right)}
\newcommand\bgv{\bigg\vert}              

\newcommand\bc{\begin{center}}
\newcommand\ec{\end{center}}

\newcommand\partder[2]{\frac{{\partial {#1}}}{{\partial {#2}}}}

\renewcommand\a{\alpha}

\renewcommand\d{\delta}
\newcommand\eps{\epsilon}
\newcommand\vareps{\varepsilon}

\newcommand\G{\Gamma}

\newcommand\h{\frac{1}{2}}
\renewcommand\k{\kappa}
\renewcommand\l{\lambda}
\renewcommand\L{\Lambda}
\newcommand\m{\mu}
\newcommand\n{\nu}

\renewcommand\O{\Omega}

\newcommand\vp{\varphi}
\renewcommand\P{\Phi}
\newcommand\pa{\partial}

\newcommand\pr{\prime}

\newcommand\s{\sigma}

\renewcommand\t{\tau}

\newcommand\cA{{\mathcal A}}
\newcommand\cB{{\mathcal B}}

\newcommand\cF{{\mathcal F}}

\newcommand\cU{{\mathcal U}}

\newcommand{\ct}[1]{\cite{#1}}
\newcommand\PRL[3]{{\em Phys. Rev. Lett.} \textbf{#1}, #3 (#2)}

\newcommand\CMP[3]{{\em Commun. Math. Phys.} \textbf{#1}, #3 (#2)}
\newcommand\PRD[3]{{\em Phys. Rev.} \textbf{D#1}, #3 (#2)}

\newcommand\PLB[3]{{\em Phys. Lett.} \textbf{#1B}, #3 (#2)}
\newcommand\CQG[3]{{\em Class. Quantum Grav.} \textbf{#1}, #3 (#2)}

\newcommand\AoP[3]{{\em Ann. of Phys.} \textbf{#1}, #3 (#2)}

\newcommand\IJMPA[3]{{\em Int. J. Mod. Phys.} \textbf{A#1}, #3 (#2)}
\newcommand\IJMPD[3]{{\em Int. J. Mod. Phys.} \textbf{D#1}, #3 (#2)}

\newcommand\MPLA[3]{{\em Mod. Phys. Lett.} \textbf{A#1}, #3 (#2)}

\makeindex

\begin{document}

\chapter[]{Confinement/Deconfinement and \\ 
Gravity-Assisted Emergent Higgs Mechanism \\
in Quintessential Cosmological Model\label{bekenstein}}

\author[]{Eduardo Guendelman}
\address{Physics Department, Ben Gurion University of the Negev,\\
Beer Sheva, Israel, 
guendel@bgu.ac.il}

\author[]{Emil Nissimov and Svetlana Pacheva}
\address{Institute for Nuclear Research and Nuclear Energy,\\
Bulgarian Academy of Sciences, Sofia, Bulgaria, \\
nissimov@inrne.bas.bg, svetlana@inrne.bas.bg}

\begin{abstract}
Motivated by the ideas of Jacob Bekenstein concerning gravity-assisted symmetry 
breaking, we consider a non-canonical model of $f(R)=R+R^2$ extended gravity 
coupled to neutral scalar ``inflaton'', as well as to $SU(2)\times U(1)$ multiplet 
of fields matching the content of the bosonic sector of the electroweak particle 
model, however with the following significant difference
-- the $SU(2)\times U(1)$ iso-doublet Higgs-like scalar enters here with a standard  
positive mass squared and without quartic selfinteraction.
Strong interaction dynamics and, in particular, QCD-like confinement effects 
are also  considered by introducing an additional coupling to a strongly
nonlinear gauge field whose Lagrangian contains a square-root of the
standard Maxwell/Yang-Mills kinetic term. The latter is known to produce
charge confinement in flat spacetime.

The principal new ingredient in the present approach is employing the
formalism of non-Riemannian spacetime volume-forms -- alternative generally covariant 
volume elements 
independent of the spacetime
metric, constructed in terms of auxiliary antisymmetric tensor gauge fields
of maximal rank. Although being almost pure-gauge, \textsl{i.e.} not introducing 
any additional propagating degrees of freedom, their dynamics triggers a series 
of physically important features when passing to the Einstein frame: 
(i) Appearance of two infinitely large flat regions of the effective
``inflaton'' scalar potential with vastly different energy scales corresponding 
to the ``early'' and ``late'' epochs of the Universe; (ii) Dynamical generation of 
Higgs-like spontaneous symmetry breaking effective potential for the 
$SU(2)\times U(1)$ iso-doublet scalar in the ``late'' Universe, and
vanishing of the symmetry breaking in the ``early'' Universe; (iii)
Dynamical appearance of charge confinement via the ``square-root'' nonlinear
gauge field in the ``late'' Universe and deconfinement in the ``early''
Universe.

\vspace{.1in}
Keywords: non-Riemannian volume-forms; quintessential evolution; confining
gauge theories, dynamical generation of electroweak symmetry breaking.

PACS numbers: 
04.50.Kd, 
11.30.Qc  
\end{abstract}


\body


\section{Introduction}
\label{intro}

Jacob Bekenstein was a remarkable scientist and person. He had both the
creativity and the courage to look at physics from a different
perspective. That was evident from his very early idea that black holes must have
entropy \ct{bekenstein-1,bekenstein-2,bekenstein-3,bekenstein-4}.
Bekenstein presented very strong arguments to support this idea but he was
confronted with the opposition of S. Hawking, who strongly argued against
his idea. The issue was later resolved in favor of Jacob when Hawking showed
that the black holes emit radiation \ct{hawking-radiation-1,hawking-radiation-2} in 
a way consistent with the entropy proposal.

Jacob had also his original approach to other fundamental physics problems, 
like the idea that gravity must be modified in order to
reproduce effects of dark matter, without the dark matter being really
present \ct{bekenstein-DM}.

Another topic Jacob was fascinated with, and on which we will focus here, was
the search for new avenues for spontaneous symmetry breaking. 
One of us (E.G.) was for example involved in a joint research with Jacob on 
how symmetry breaking can be generated by density effects, even for a theory 
without spontaneous symmetry breaking when no
background densities were present \ct{bekenstein-eduardo-87}.
During this collaboration E.G. learned lots of physics, of course, but also
he understood how it is possible for a great mind to be simultaneously a remarkable
human being.

Bekenstein's search for new mechanisms for symmetry breaking lead him also to
consider gravity-induced symmetry breaking effects instead of density
effects. In an intriguing paper \ct{gravity-assist-86} from 1986 he proposed the
remarkable idea of a gravity-assisted spontaneous symmetry breaking
of electroweak (Higgs) type without invoking unnatural (according to his 
opinion) ingredients like negative mass squared and a quartic
self-interaction for the Higgs field. By considering a model of gravity
interacting with a standard Klein-Gordon scalar field (with small positive mass
squared and without selfinteraction) coupled conformally to the scalar
curvature he managed to obtain a prototype of dynamically induced
Higgs-like spontaneous symmetry breaking scalar potential. A similar
approach was further worked out in Ref.~\refcite{moniz-etal}.  

Motivated by Bekenstein's idea, we wrote an essay \ct{grf-essay}
to the gravity research foundation, where we considered a
non-canonical model of gravity coupled to a neutral scalar ``inflaton'' 
as well as to a set of $SU(2)\times U(1)$ iso-doublet scalar and gauge fields 
corresponding to the bosonic sector of the electroweak particle model. Here
the iso-doublet scalar field was introduced with a standard positive mass squared
and without selfinteraction.

The essential non-standard feature of the model in Ref.~\refcite{grf-essay} is
its construction in terms of non-Riemannian spacetime volume-forms
(alternative metric-independent generally covariant volume elements) defined in
terms of auxiliary antisymmetric tensor gauge fields of maximal rank.
The latter were shown to be almost pure-gauge -- apart from few arbitrary 
integration constants they do not produce propagating field-theoretic degrees of 
freedom (see Appendices A of Refs.~\refcite{grav-bags,grf-essay}).
Yet the non-Riemannian spacetime volume-forms trigger a series of important 
physical features unavailable in ordinary gravity-matter models with the 
standard Riemannian volume-form (given by the square-root of the determinant 
of the Riemannian metric):

(i) The ``inflaton'' $\vp$ develops a remarkable effective scalar potential in
the Einstein frame possessing an infinitely large flat region for large
negative $\vp$ describing the ``early'' universe evolution; 

(ii) In the absence of the $SU(2)\times U(1)$ iso-doublet scalar field, 
the ``inflaton'' effective potential has another infinitely large flat region 
for large positive $\vp$ at much lower energy scale describing the ``late'' 
post-inflationary (dark energy dominated) universe;

(iii) Inclusion of the $SU(2)\times U(1)$ iso-doublet scalar field $\s$ introduces
a drastic change in the total effective scalar potential in the
post-inflationary universe -- the effective potential as a function of $\s$ 
dynamically acquires exactly the electroweak Higgs-type spontaneous symmetry 
breaking form.

In the present paper we will extend the above model by introducing in the
initial action a $R^2$-gravity term as well as coupling to an additional
strongly nonlinear gauge field whose Lagrangian contains a square-root of
the standard Maxwell/Yang-Mills kinetic term. The latter is known to
describe charge confinement in flat spacetime \ct{GG-2} as well as 
in curved spacetime for static spherically symmetric field configurations 
(Appendix B in Ref.~\refcite{grav-bags}; see also Eq.\rf{cornell-type} below). Thus, 
the addition of the ``square-root'' nonlinear gauge field will simulate the strong
interactions QCD-like dynamics and, therefore, our extended model represents
qualitatively a quintessential cosmological model incorporating the full bosonic
content of the standard particle model. Now, in the physical Einstein frame 
alongside with the Bekenstein-inspired gravity-assisted dynamical generation 
of Higgs-type electroweak spontaneous symmetry breaking in the ``late'' 
universe, while there is no electroweak breaking in the ``early'' universe, we 
obtain gravity-assisted dynamical generation of charge confinement in the
``late'' universe as well as gravity-suppression of confinement,
\textsl{i.e.}, deconfinement in the ``early'' universe.

\section{Non-Canonical Gravity Coupled to a Confining Nonlinear Gauge Field
and the Bosonic Sector of the Electroweak Standard Model}
\label{noncanon-grav}
\subsection{Non-Standard $f(R)$-Gravity Model with Non-Riemannian Spacetime
Volume-Forms}
\label{TMMT}

We start with the following non-canonical $f(R)=R+R^2$ gravity-matter action 
constructed in terms of two different non-Riemannian volume-forms (generally
covariant metric-independent volume elements) generalizing the actions in
Refs.~\refcite{grav-bags,grf-essay} (for simplicity we use units with the Newton
constant $G_N = 1/16\pi$):
\br
S = \int d^4 x\,\P (A) \Bigl\lb R + L_1 (\vp,X) + L_2 (\s,Y) 
- \h f_0 \sqrt{-F^2}\Bigr\rb +
\nonu \\
\int d^4 x\,\P (B) \Bigl\lb \eps R^2 - \frac{1}{4e^2} F^2  
- \frac{1}{4g^2} \cF^2(\cA) - \frac{1}{4g^{\pr\,2}} \cF^2(\cB) +
\frac{\P (H)}{\sqrt{-g}}\Bigr\rb  \; .
\lab{TMMT-1}
\er
Here the following notations are used:
\begin{itemize}
\item
$\P(A)$ and $\P (B)$ are two independent non-Riemannian volume-forms given in 
terms of the dual field-strengths of rank 3 antisymmetric tensor gauge fields 
$A_{\n\k\l}$ and $B_{\n\k\l}$.
\be
\P (A) = \frac{1}{3!}\vareps^{\m\n\k\l} \pa_\m A_{\n\k\l} \quad ,\quad
\P (B) = \frac{1}{3!}\vareps^{\m\n\k\l} \pa_\m B_{\n\k\l} \; .
\lab{Phi-1-2}
\ee
\item
$\P (H)$ is the dual field-strength of an additional auxiliary tensor gauge field 
$H_{\n\k\l}$, whose presence is crucial for the consistency of the model 
\rf{TMMT-1}:
\be
\P (H) = \frac{1}{3!}\vareps^{\m\n\k\l} \pa_\m H_{\n\k\l} \; .
\lab{Phi-H}
\ee
\item
We particularly emphasize that we start within the first-order 
{\em Palatini formalism} for the scalar curvature $R$ and the Ricci tensor
$R_{\m\n}$: $R=g^{\m\n} R_{\m\n}(\G)$, where $g_{\m\n}$, $\G^\l_{\m\n}$ --
the metric and affine connection are {\em apriori} independent.
\item
$L_1 (\vp,X)$ is the ``inflaton'' Lagrangian:
\be
L_1 (\vp,X) = X - f_1 e^{-\a\vp} \;\; ,\;\;
X \equiv - \h g^{\m\n} \pa_\m \vp \pa_\n \vp \; ,
\lab{L-1}
\ee
where $\a, f_1$ are dimensionful positive parameters.
\item
$\s \equiv (\s_a)$ is a complex $SU(2)\times U(1)$ iso-doublet Higgs-like 
scalar field with Lagrangian:
\be
L_2 (\s,Y) = Y - m_0^2 \s^{*}_a \s_a \;\; ,\;\; 
Y \equiv - g^{\m\n}(\nabla_\m \s)^{*}_a \nabla_\n \s_a \; ,
\lab{L-2}
\ee
where gauge-covariant derivative acting on $\s$ reads:
\be
\nabla_\m \s = 
\Bigl(\pa_\m - \frac{i}{2} \t_A \cA_\m^A - \frac{i}{2} \cB_\m \Bigr)\s \; ,
\lab{cov-der}
\ee
with $\h \t_A$ ($\t_A$ -- Pauli matrices, $A=1,2,3$) indicating the $SU(2)$ 
generators and $\cA_\m^A$ ($A=1,2,3$)
and $\cB_\m$ denoting the corresponding electroweak $SU(2)$ and $U(1)$ gauge fields.
\item
The electroweak gauge field kinetic terms are of the standard Yang-Mills form 
(all $SU(2)$ indices $A,B,C = (1,2,3)$):
\br
\cF^2(\cA) \equiv \cF^A_{\m\n} (\cA) \cF^A_{\k\l} (\cA) g^{\m\k} g^{\n\l} \;\; ,\;\;
\cF^2(\cB) \equiv \cF_{\m\n} (\cB) \cF_{\k\l} (\cB) g^{\m\k} g^{\n\l} \; ,
\lab{F2-def} \\
\cF^A_{\m\n} (\cA) = 
\pa_\m \cA^A_\n - \pa_\n \cA^A_\m + \eps^{ABC} \cA^B_\m \cA^C_\n \;\; ,\;\;
\cF_{\m\n} (\cB) = \pa_\m \cB_\n - \pa_\n \cB_\m \; .
\lab{F-def}
\er
\end{itemize}

Finally, there is an additional coupling in the action \rf{TMMT-1} to another 
strongly nonlinear (Abelian) gauge field $A_\m$ with the square-root Maxwell term 
$-\h f_0\sqrt{-F^2}$ alongside the standard kinetic term $-\frac{1}{4e^2} F^2$:
\be
F^2 \equiv F_{\m\n}F_{\k\l}g^{\m\k}g^{\n\l} \;\; ,\;\; 
F_{\m\n} = \pa_\m A_\n - \pa_\n A_\m \; .
\lab{confining}
\ee
As shown in Appendix B of Ref.~\refcite{grav-bags}, for static spherically 
symmetric fields in a static spherically symmetric spacetime metric the 
square-root term $-\h f_0\sqrt{-F^2}$ produces an effective 
{\em ``Cornell''-type confining potential}
\ct{cornell-potential-1,cornell-potential-2,cornell-potential-3}
$V_{\rm eff}(L)$ between charged quantized fermions, $L$ being the distance 
between the latter:
\be
V_{\rm eff} (L) = \sqrt{2}ef_0\; L - \frac{e^2}{2\pi\,L} + 
\bigl( L{\rm -independent} ~{\rm const} \bigr) \; ,
\lab{cornell-type}
\ee
\textsl{i.e.}, $f_0$ and $e$ play the role of a confinement-strength coupling 
constant and of a ``color'' charge, respectively.

In fact, we could equally well take the ``square-root'' nonlinear gauge field 
$A_\m$ to be non-Abelian -- for static spherically symmetric solutions 
the non-Abelian model effectively reduces to the abelian one \ct{GG-2}.
Thus, the ``square-root'' gauge field will simulate the QCD-like confining
dynamics.

Let us note that the structure of action \rf{TMMT-1} is uniquely fixed by the 
requirement for invariance (with the exception of the regular mass term 
of the iso-doublet scalar $\s_a$) under the following global Weyl-scale 
transformations:
\br
g_{\m\n} \to \l g_{\m\n} \;\; ,\;\; \vp \to \vp + \frac{1}{\a}\ln \l \;\;,\;\; 
A_{\m\n\k} \to \l A_{\m\n\k} \;\; ,\;\; B_{\m\n\k} \to \l^2 B_{\m\n\k} \; ,
\lab{scale-transf} \\
\G^\m_{\n\l} \;,\; H_{\m\n\k} \;,\; \s_a \; ,\; A_\m \; ,\;\cA^A_\m \; ,\; \cB_\m \;
-- \; {\rm inert} \; .
\nonu
\er

\subsection{Derivation of the Einstein-Frame Action}
\label{einstein-frame-derivation}

Solutions of the equations of motion of the initial action \rf{TMMT-1} w.r.t. 
auxiliary tensor gauge fields $A_{\m\n\l}$, $B_{\m\n\l}$ and $H_{\m\n\l}$ 
yield the following algebraic constraints:
\br
R + L_1 (\vp,X) + L_2 (\s,Y) -\h f_0 \sqrt{-F^2} = - M_1 = {\rm const} \; ,
\lab{integr-const-1} \\
\eps R^2 - \frac{1}{4e^2} F^2 - \frac{1}{4g^2} \cF^2(\cA) 
- \frac{1}{4g^{\pr\,2}} \cF^2(\cB) + \frac{\P (H)}{\sqrt{-g}}
= - M_2  = {\rm const} \; ,
\lab{integr-const-2}\\
\frac{\P(B)}{\sqrt{-g}} \equiv \chi_2 = {\rm const} \; ,
\lab{integr-const-3}
\er
where $M_1$ and $M_2$ are arbitrary dimensionful and $\chi_2$
arbitrary dimensionless {\em integration constants}. The algebraic constraint
Eqs.\rf{integr-const-1}-\rf{integr-const-3} are the Lagrangian-formalism counterparts of the
Dirac first-class Hamiltonian constraints on the auxiliary tensor gauge fields
$A_{\m\n\l},\, B_{\m\n\l},\, H_{\m\n\l}$ \ct{grav-bags,grf-essay}.

The equations of motion of \rf{TMMT-1} w.r.t. affine connection $\G^\m_{\n\l}$ 
(recall -- we are using Palatini formalism):
\be
\int d^4\,x\,\sqrt{-g} g^{\m\n} \Bigl(\frac{\P_1}{\sqrt{-g}} +
2\eps\,\frac{\P_2}{\sqrt{-g}}\, R\Bigr) \(\nabla_\k \d\G^\k_{\m\n}
- \nabla_\m \d\G^\k_{\k\n}\) = 0 
\lab{var-G}
\ee
yield a solution for $\G^\m_{\n\l}$ as a Levi-Civita connection:
\be
\G^\m_{\n\l} = \G^\m_{\n\l}({\bar g}) = 
\h {\bar g}^{\m\k}\(\pa_\n {\bar g}_{\l\k} + \pa_\l {\bar g}_{\n\k} 
- \pa_\k {\bar g}_{\n\l}\) \; ,
\lab{G-eq}
\ee
w.r.t. to the following {\em Weyl-rescaled metric} ${\bar g}_{\m\n}$:
\be
{\bar g}_{\m\n} = \bigl(\chi_1 + 2\eps\chi_2 R\bigr) g_{\m\n} 
\quad , \quad
\chi_1 \equiv \frac{\P_1 (A)}{\sqrt{-g}} \; ,
\lab{bar-g}
\ee
$\chi_2$ as in \rf{integr-const-3}.
Upon using relation \rf{integr-const-1} and notation \rf{integr-const-3}
Eq.\rf{bar-g} can be written as:
\be
{\bar g}_{\m\n} = \Bigl\lb\chi_1 - 2\eps\chi_2 \Bigl(L_1(\vp,X) + L_2(\s,Y) 
-\h f_0\sqrt{-F^2} + M_1\Bigr)\Bigr\rb g_{\m\n}\; .
\lab{bar-g-1}
\ee

Varying \rf{TMMT-1} w.r.t. the original metric $g_{\m\n}$ and using relations 
\rf{integr-const-1}-\rf{integr-const-3} we have:
\be
\chi_1 \Bigl\lb R_{\m\n} + \h\( g_{\m\n}L^{(1)} - T^{(1)}_{\m\n}\)\Bigr\rb -
\h \chi_2 \Bigl\lb T^{(2)}_{\m\n} + g_{\m\n} \(\eps R^2 + M_2\)
- 4\eps R\,R_{\m\n}\Bigr\rb = 0 \; ,
\lab{pre-einstein-eqs}
\ee
with $\chi_1$ and $\chi_2$ as in \rf{bar-g} and \rf{integr-const-3},
and $T^{(1,2)}_{\m\n}$ being the canonical energy-momentum tensors:
\be
T^{(1,2)}_{\m\n} = g_{\m\n} L^{(1,2)} - 2 \partder{}{g^{\m\n}} L^{(1,2)} \; .
\lab{EM-tensor}
\ee
of the scalar+gauge field Lagrangians in the original action \rf{TMMT-1}:
\be
L^{(1)} \equiv L_1 (\vp,X) + L_2 (\s,Y) -\h f_0 \sqrt{-F^2} \;\; ,\;\;
L^{(2)} \equiv - \frac{1}{4e^2} F^2 - \frac{1}{4g^2} \cF^2(\cA) 
- \frac{1}{4g^{\pr\,2}} \cF^2(\cB) \; .
\lab{L-1-2-def}
\ee

Taking the trace of Eqs.\rf{pre-einstein-eqs} and using again relation 
\rf{integr-const-1} we solve for the ratio $\chi_1$ \rf{bar-g}:
\be
\chi_1 = 2 \chi_2 \frac{T^{(2)}/4 + M_2}{L^{(1)} - \h T^{(1)} - M_1} \; ,
\lab{chi-1-eq}
\ee
where $T^{(1,2)} = g^{\m\n} T^{(1,2)}_{\m\n}$. Explicitly we obtain from 
\rf{chi-1-eq}:
\be
\chi_1 = \frac{1}{2\chi_2 M_2} \Bigl( f_1 e^{-\a\vp} + m_0 \s^{*}\s - M_1 \Bigr)
\lab{chi-1-sol}
\ee

The Weyl-rescaled metric ${\bar g}_{\m\n}$ \rf{bar-g-1} can be written
explicitly as:
\br
{\bar g}_{\m\n} = \chi_1 \O g_{\m\n} \;\; ,\;\;
\O \equiv \frac{1+\frac{\eps}{M_2}\bigl( 
f_1 e^{-\a\vp} + m_0 \s^{*}\s - M_1\bigr)^2}{1+2\eps\chi_2 \bigl({\bar X} +
{\bar Y} - \h f_0 \sqrt{-{\bar F}^2}\bigr)} \; ,
\lab{bar-g-2} \\
{\bar X} \equiv - \h {\bar g}^{\m\n} \pa_\m \vp \pa_\n \vp \;\; ,\;\;
{\bar Y} \equiv - {\bar g}^{\m\n}(\nabla_\m \s)^{*}_a \nabla_\n \s_a \;\;,
\;\; {\bar F}^2 \equiv F_{\m\n}F_{\k\l} {\bar g}^{\m\k} {\bar g}^{\n\l} \; .
\lab{X-Y-F-bar}
\er

Now, we can bring Eqs.\rf{pre-einstein-eqs} into the standard form of Einstein 
equations in the second-order formalism for the Weyl-rescaled  metric
${\bar g}_{\m\n}$ \rf{bar-g-2}, \textsl{i.e.}, the {\em Einstein-frame} equations: 
\be
R_{\m\n}({\bar g}) - \h {\bar g}_{\m\n} R({\bar g}) = \h T^{\rm eff}_{\m\n}
\lab{eff-einstein-eqs}
\ee
with effective energy-momentum tensor corresponding according to the definition 
\rf{EM-tensor}:
\be
T^{\rm eff}_{\m\n} = g_{\m\n} L_{\rm eff} - 2 \partder{}{g^{\m\n}} L_{\rm eff}
\lab{T-eff}
\ee
to the following effective {\em Einstein-frame} matter Lagrangian (using
short-hand notations \rf{L-1-2-def}) and with $\chi_1$ as in \rf{chi-1-sol}
and $\O$ as in \rf{bar-g-2}):
\be
L_{\rm eff} = \frac{1}{\chi_1\O}\Bigl\{ L^{(1)} + M_1 +
\frac{\chi_2}{\chi_1\O}\Bigl\lb L^{(2)} + M_2 + 
\eps (L^{(1)} + M_1)^2\Bigr\rb\Bigr\} \; .
\lab{L-eff}
\ee

The full Einstein-frame action, where all quantities defined w.r.t.
Einstein-frame metric \rf{bar-g} are indicated by an upper bar, explicitly reads:
\be
S = \int d^4 x \sqrt{-{\bar g}} \Bigl\lb R({\bar g}) +
L_{\rm eff} \bigl(\vp,{\bar X};\s,{\bar Y}; {\bar F}^2, 
{\bar\cF(\cA)}^2,{\bar\cF(\cB)}^2\bigr)\Bigr\rb \; ,
\lab{TMMT-einstein-frame}
\ee
where ${\bar X},\, {\bar Y},\, {\bar F}^2$ are as in \rf{X-Y-F-bar} (and
similarly for ${\bar\cF(\cA)}^2,\, {\bar\cF(\cB)}^2$), and where:
\br
L_{\rm eff} = \bigl({\bar X}+{\bar Y}\bigr)\bigl(1-4\eps\chi_2 \cU(\vp,\s)\bigr) +
\eps\chi_2 \bigl({\bar X}+{\bar Y}\bigr)^2 \bigl(1-4\eps\chi_2 \cU(\vp,\s)\bigr)
\nonu \\
- \bigl({\bar X}+{\bar Y}\bigr) \sqrt{-{\bar F}^2} \eps\chi_2\, f_{\rm eff}(\vp,\s) 
- \h f_{\rm eff}(\vp,\s) \sqrt{-{\bar F}^2}
\nonu \\
- \cU(\vp,\s) - \frac{1}{4 e^2_{\rm eff}(\vp,\s)}{\bar F}^2 
-\frac{\chi_2}{4g^2}{\bar\cF}^2(\cA) -\frac{\chi_2}{4g^{\pr\,2}}{\bar\cF}^2(\cB)
\; 
\lab{L-eff-total}
\er
In \rf{L-eff-total} the following notations are used:
\begin{itemize}
\item
$\cU(\vp,\s)$ is the effective scalar field (``inflaton'' + Higgs-like) potential:
\be
\cU(\vp,\s) = \frac{\bigl( f_1 e^{-\a\vp} + m_0 \s^{*}\s - M_1\bigr)^2}{4\chi_2
\bigl\lb M_2 + \eps \bigl( f_1 e^{-\a\vp} + m_0 \s^{*}\s - M_1\bigr)^2\bigr\rb} \; .
\lab{U-vp-s}
\ee
\item
$f_{\rm eff}(\vp,\s)$ is the effective confinement-strength coupling constant:
\be
f_{\rm eff}(\vp,\s) = f_0 \bigl(1-4\eps\chi_2 \cU(\vp,\s)\bigr) \; ;
\lab{f-eff}
\ee
\item
$e^2_{\rm eff}(\vp,\s)$ is the effective ``color'' charge squared:
\be
e^2_{\rm eff}(\vp,\s) = \frac{e^2}{\chi_2}
\Bigl\lb 1 + \eps e^2 f_0^2 \bigl(1-4\eps\chi_2 \cU(\vp,\s)\bigr) \Bigr\rb^{-1}
\lab{e-eff}
\ee
\end{itemize}

Note that \rf{L-eff-total} is of quadratic {\em ``k-essence''} type
\ct{k-essence-1,k-essence-2,k-essence-3,k-essence-4} w.r.t.
``inflaton'' $\vp$ and the Higgs-like $\s$ fields.


\section{Quintessence, Confinement/Deconfinement and Gravity Assisted
Emergent Higgs Mechanism}
\label{quintess}

The nonlinear ``confining'' gauge field $A_\m$ develops a nontrivial vacuum
field-strength:
\be
\frac{\pa L_{\rm eff}}{\pa {\bar F}^2}\bgv_{{\bar X},{\bar Y}=0} = 0
\lab{F-vac-eq}
\ee
explicitly given by:
\be
\sqrt{-{\bar F}^2}_{\rm vac} = f_{\rm eff}(\vp,\s)\, e^2_{\rm eff}(\vp,\s)
\lab{F-vac}
\ee
Substituting \rf{F-vac} into \rf{L-eff-total} we obtain the following total
effective scalar potential (with $\cU(\vp,\s)$ as in \rf{U-vp-s}):
\be
\cU_{\rm total}(\vp,\s) = \frac{\cU(\vp,\s)(1-\eps e^2 f_0^2) + e^2 f_0^2/4\chi_2}{1
+ \eps e^2 f_0^2 \bigl(1-4\eps\chi_2 \cU(\vp,\s)\bigr)} \; .
\lab{U-eff-total}
\ee

$\cU_{\rm total}(\vp,\s)$ \rf{U-eff-total} has few remarkable properties.

First, $\cU_{\rm total}(\vp,\s)$ possesses two infinitely large flat regions
as function of $\vp$ when $\s$ is fixed:

(a) (-) flat ``inflaton'' region for large negative values of $\vp$,

(b) (+) flat ``inflaton'' region for large positive values of $\vp$,

respectively, as depicted on Fig.1 (for $m_0 \s^{*}\s \leq M_1$) or 
Fig.2 (for $m_0 \s^{*}\s \geq M_1$).

\begin{figure}
\begin{center}
\includegraphics[width=9cm,keepaspectratio=true]{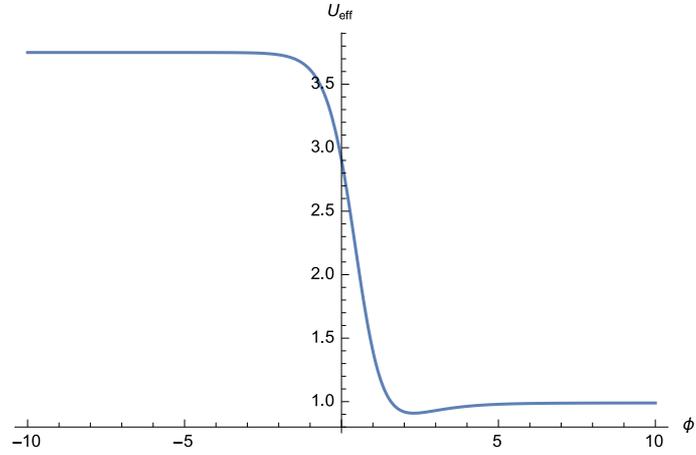}
\caption{Qualitative shape of the total effective scalar potential $U_{\rm total}$ 
\rf{U-eff-total} as function of the ``inflaton''$\vp$ for fixed Higgs-like $\s$
(when $m_0 \s^{*}\s \leq M_1$).}
\end{center}
\end{figure}

\begin{figure}
\begin{center}
\includegraphics[width=9cm,keepaspectratio=true]{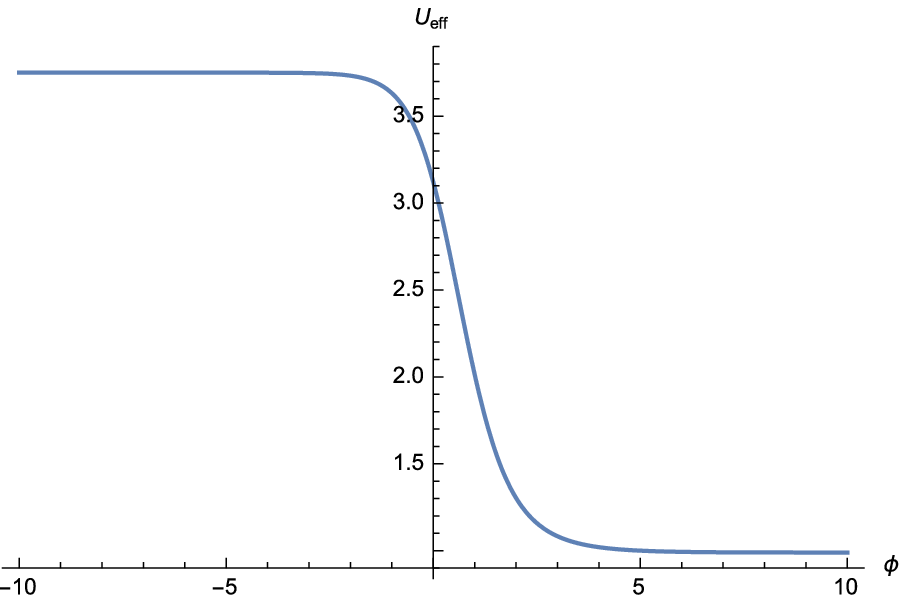}
\caption{Qualitative shape of the total effective scalar potential $U_{\rm total}$ 
\rf{U-eff-total} as function of the ``inflaton''$\vp$ for fixed Higgs-like $\s$
(when $m_0 \s^{*}\s \geq M_1$).}
\end{center}
\end{figure}

\vspace{0.1in}

(i) In the (-) flat ``inflaton'' region:
\begin{itemize}
\item
The effective scalar field potential reduces to:
\be
\cU(\vp,\s ={\rm fixed}) \simeq \frac{1}{4\eps\chi_2}\quad\longrightarrow\quad
\cU_{\rm total} \simeq \cU^{(-)}_{\rm total} = \frac{1}{4\eps\chi_2} \; ,
\lab{U-minus}
\ee
implying that all terms containing $\vp$ and $\s$ disappear from the
Einstein-frame Lagrangian \rf{TMMT-einstein-frame}, \textsl{i.e.}, there is
{\em no electroweak spontaneous breakdown in the (-) flat ``inflaton'' region}. 
\item
From \rf{f-eff} the first relation \rf{U-minus} implies $f_{\rm eff} = 0$, 
\textsl{i.e.}, there is {\em no confinement in the (-) flat ``inflaton'' region}.
\end{itemize}

\vspace{0.1in}

(ii) In the (+) flat ``inflaton'' region:
\begin{itemize}
\item
The effective scalar field potential becomes:
\br
\cU(\vp,\s) \simeq \cU_{(+)}(\s) = \frac{\bigl(m_0^2 \s^{*}\s - M_1\bigr)^2}{
4\chi_2 \bigl\lb M_2 + \eps \bigl(m_0^2 \s^{*}\s - M_1\bigr)^2\bigr\rb}
\lab{U-plus} \\
\longrightarrow \quad 
\cU_{\rm total} (\vp\s) \simeq \cU^{(+)}_{\rm total}(\s) = 
\frac{\cU_{(+)}(\s)(1-\eps e^2 f_0^2) + e^2 f_0^2/4\chi_2}{1
+ \eps e^2 f_0^2 \bigl(1-4\eps\chi_2 \cU_{(+)}(\s)\bigr)}
\lab{U-total-plus}
\er
producing a dynamically generated {\em nontrivial vacuum for the Higgs-like field}:
\be
|\s_{\rm vac}|= \sqrt{M_1}/m_0 \; ,
\lab{higgs-vac}
\ee
\textsl{i.e.}, we obtain {\em ``gravity-assisted'' electroweak spontaneous breakdown
in the (+) flat ``inflaton'' region}.
\item
At the Higgs vacuum we have dynamically generated vacuum energy density
(cosmological constant):
\be
\cU^{(+)}_{\rm total}(\s_{\rm vac}) \equiv 2 \L_{(+)} = \eps e^2 f_0^2
\Bigl\lb 4\eps\chi_2 \bigl( 1 + \eps e^2 f_0^2\bigr)\Bigr\rb^{-1} \; .
\lab{CC-plus}
\ee
\item
The effective confinement-strength coupling constant:
\be
f_{\rm eff} \simeq f_{(+)} = f_0 \bigl(1-4\eps\chi_2 \cU_{(+)}(\s)\bigr) > 0\; ,
\lab{f-plus}
\ee
threfore we obtain {\em ``gravity-assisted'' charge confinement in the 
(+) flat ``inflaton'' region}.
\end{itemize}

As seen from Fig.1 or Fig.2, the two flat ``inflaton'' regions of the total scalar 
potential given by $\cU^{(-)}_{\rm total} = \frac{1}{4\eps\chi_2}$ \rf{U-minus} 
and $\cU^{(+)}_{\rm total}(\s_{\rm vac}) \equiv 2 \L_{(+)} = \eps e^2 f_0^2
\Bigl\lb 4\eps\chi_2 \bigl( 1 + \eps e^2 f_0^2\bigr)\Bigr\rb^{-1}$
\rf{CC-plus}, respectively, can be identified as describing the
``early'' (``inflationary'') and ``late'' (today's dark energy dominated) epoch of 
the universe provided we take the following numerical values for the parameters 
in order to conform to the {\em PLANCK} data \ct{Planck-1,Planck-2}:
\be
\cU^{(-)}_{\rm total} \sim 10^{-8} M_{\rm Pl}^4 \to 
\eps\chi_2 \sim 10^8 M_{\rm Pl}^{-4}
\;\; ,\;\; \L_{(+)} \sim 10^{-122} M_{\rm Pl}^4 \to 
\frac{e^2 f_0^2}{\chi_2} \sim 10^{-122} M_{\rm Pl}^4 \; ,
\lab{param-1}
\ee
where $M_{\rm Pl}$ is the Planck mass scale.

From the Higgs v.e.v. $|\s_{\rm vac}|= \sqrt{M_1}/m_0$ and the Higgs mass
$\frac{M_1 m_0^2}{4\chi_2 M_2}$ resulting from the dynamically generated
Higgs-like potential $\cU^{(+)}_{\rm total}(\s)$ \rf{U-total-plus} we find:
\be
m_0 \sim M_{\rm EW} \;\; ,\;\; M_{1,2}\sim M_{\rm EW}^4 \; ,
\lab{param-2}
\ee
where $M_{\rm EW} \sim 10^{-16} M_{\rm Pl}$ is the electroweak mass scale.




\section{Conclusions and Outlook}
\label{conclude}

Here we have proposed a non-canonical model of $f(R)=R+R^2$ gravity coupled to 
non-standard  matter incorporating two main building blocks -- employing the 
formalism of non-Riemannian spacetime volume forms (generally covariant
metric-independent volume elements) as well as introducing a special strongly 
non-linear gauge field with a square-root of the usual Maxwell/Yang-Mills
kinetic term simulating QCD-like confinement dynamics. 
Due to the special interplay of the dynamics of the above principal
ingredients our model is capable of producing in the Einstein frame:

\begin{itemize}
\item
(i) Unified ``quintessential'' description of the evolution of the ``early''
and ``late'' universe due to a natural dynamical generation of vastly
different vacuum energy densities thanks to the auxiliary non-Riemannian
volume-form antisymmetric tensor gauge fields;
\item
(ii) Gravity-assisted dynamical generation of Higgs-like electroweak spontaneous
symmetry breaking effective scalar potential in the ``late'' universe, as
well as gravity-assisted charge confinement mechanism through the
``square-root'' nonlinear gauge field;
\item
(iii) Gravity-induced suppression of electroweak spontaneous symmetry breaking, as
well as gravity-induced deconfinement in the ``early'' universe.
\end{itemize}

The non-Riemannian volume-form formalism has further physically relevant
applications such as producing a novel mechanism for supersymmetric
Brout-Englert-Higgs effect in supergravity through dynamical generation of a
cosmological constant triggering spontaneous supersymmetry breaking and
dynamical gravitino mass generation \ct{susyssb-1,susyssb-2}.

Similarly, the QCD-simulating ``square-root'' nonlinear gauge field when
interacting with gravity produces several other interesting effects:

\begin{itemize}
\item
(a) black holes with an additional constant
background electric field exercising confining force on charged test
particles even when the black hole itself is electrically neutral
\ct{grav-cornell}; 
\item
(b) Coupling to a charged lightlike brane produces a charge-``hiding'' lightlike
thin-shell wormhole, where a genuinely charged matter source is detected as 
electrically neutral by an external observer \ct{hide-confine}.
\item
(c) Coupling to two oppositely charged lightlike brane sources produces a 
two-``throat'' lightlike thin-shell wormhole displaying a genuine QCD-like 
charge confinement, \textsl{i.e.}, the whole electric flux is trapped within
a tube-like spacetime region connected the two charged lightlike branes
\ct{hide-confine}.
\item
(d) Charge confining gravitational electrovacuum shock wave \ct{shock}.
\end{itemize}

The model here presented needs further amendments in order to avoid getting 
an unnaturally small value for the effective confinement strength coupling 
constant $f_0$ in the ``late'' universe resulting from the second relation 
\rf{param-1} (condition for compatibility with the {\em PLANCK} data 
\ct{Planck-1,Planck-2} for the value of today's cosmological constant). 

Further obvious extension of the present model must be inclusion of the
fermions in order to incorporate more faithfully the full standard particle model.
To this end we can follow the steps outlined in several previous papers by
some of us devoted to the study of modified gravity within the
non-Riemannian volume element formalism coupled to fermionic matter fields,
such as Ref.~\refcite{fermion-families} (on the geometric origins of
fermionic families), Ref.~\refcite{DE-DM-fermions} (fermionic families and
dark energy and dark matter), Ref.~\refcite{neutrino-DE} (exotic low density
fermionic states and neutrino dark energy).




\section*{Acknowledgments}
We gratefully acknowledge support of our collaboration through 
the academic exchange agreement between the Ben-Gurion University in Beer-Sheva,
Israel, and the Bulgarian Academy of Sciences. 
E.N. and E.G. have received partial support from European COST actions
MP-1405 and CA-16104, and from CA-15117 and CA-16104, respectively.
E.N. and S.P. are also thankful to Bulgarian National Science Fund for
support via research grant DN-18/1. 



\end{document}